\documentclass[aps,preprint]{revtex4}

\usepackage{graphicx}

\usepackage{keyval}

\usepackage{amsmath}

\begin{document}

\preprint{ERAU-PHY-0302}

\title{Oscillations and Cross Sections at the SNS with a 
       Large \v{C}erenkov Detector}

\author{Gordon J.\ VanDalen}

\email{vanda029@erau.edu}

\affiliation{Department of Physics, Embry-Riddle Aeronautical
University, Prescott, Arizona 86301}

\date{\today}

\begin{abstract}
MiniBooNE at FermiLab should be able to confirm or refute the LSND 
Decay-in-Flight $\nu_\mu \to \nu_e$ oscillation signal within a few years.
The primary evidence of neutrino oscillations from the LSND was in the
anti-neutrino channel $\overline{\nu}_\mu \to \overline{\nu}_e$ channel,
which may not be accessible at MiniBooNE for many years.  
The rates of signal and background are presented for a MiniBooNE style 
detector with a 250 ton mineral oil fiducial mass 
($\sim$300m$^3$ fiducial volume) placed 60 meters from the Oak Ridge SNS beam stop.  
Several hundred $\overline{\nu}_\mu \to \overline{\nu}_e$ 
events could be measured in one beam year, even at the conservative 
end of the combined analysis of LSND and KARMEN.  
The same detector could easily measure neutrino-nucleus cross sections if
filled with any interesting transparent fluid, several of which are suggested here.  
The rate and backgrounds for a methylene iodide filled detector are also presented
as an example.

\vskip 0.2in
\noindent
{\footnotesize\em Talk at Neutrino Studies at the Spallation Neutron Source
Workshop, August 28-29, 2003, Oak Ridge, TN.}
\end{abstract}

\maketitle

\section{\label{sec:intro} Introduction}

The LSND observation of anti-neutrino oscillations in the
$\overline{\nu}_\mu \to \overline{\nu}_e$ channel~\cite{LSND96} 
remains to be confirmed or refuted by an independent measurement.  
The LSND group also observed a statistically less robust signal for neutrino 
oscillations in a search for $\nu_\mu \to \nu_e$ candidates~\cite{LSNDion}.

The MiniBooNE group at FermiLab is presently collecting data to test the LSND neutrino
oscillation channel  $\nu_\mu \to \nu_e$ using the $\nu_\mu$ flux from the decay-in-flight 
(DIF) of a horn focused $\pi^+$ beam produced by 8 GeV protons~\cite{MiniBooNE}.
Collection of the positive beam (neutrino) sample for their blind analysis
will certainly take through 2004.
The limited proton beam resources at FermiLab will be redirected in early 
2005~\cite{cutoff} to other neutrino experiments~\cite{minos}.  
This may not leave MiniBooNE enough beam to gather
sufficient $\nu_\mu$ data and still be able to switch horn polarity and collect data with
the lower flux $\pi^-$ DIF source of $\overline{\nu}_\mu$.  
MiniBooNE may end in the unsatisfactory
condition of being unable to reach a strong conclusion about the primary LSND 
oscillation result in $\overline{\nu}_\mu \to \overline{\nu}_e$.
Furthermore, the MiniBooNE signal to background ratio is expected to be less than 
one, even for for the higher flux $\pi^+$ DIF beam.  The signal rate and quality are 
expected to be even less for the $\pi^-$ DIF beam, should that configuration be run.

\begin{figure}[htbp]
\includegraphics[width=0.650\textwidth]{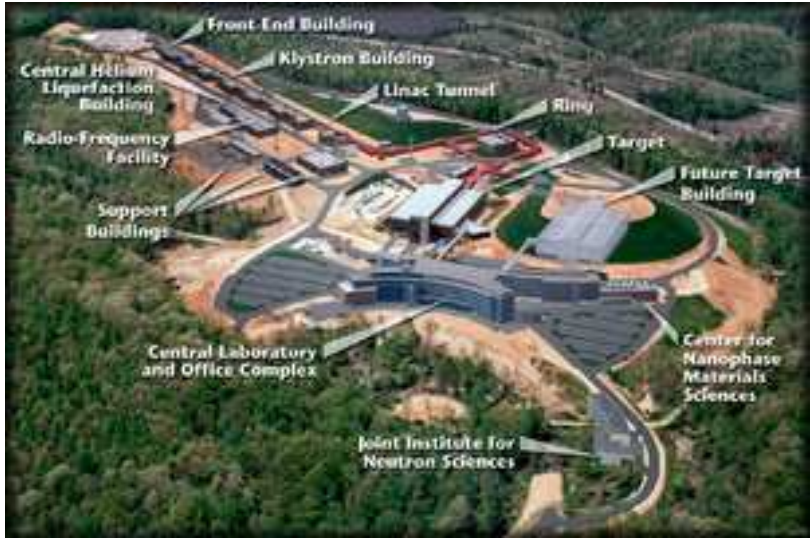}
\caption{\label{fig:SNSsite} The SNS Site.}
\end{figure}

The Spallation Neutron Source (SNS) is presently under construction at 
Oak Ridge National Lab.~\cite{SNSlab}
The SNS beam stop will provide a copious flux of decay-at-rest (DAR) neutrinos, primarily
from $\pi^+$ and $\mu^+$ decays.  
This neutrino source offers three major improvements over 
previous beam dump DAR neutrino sources.  
First, there will be a dedicated proton beam 
averaging 1.4 MW power operating throughout the year yielding a high total neutrino flux.
Second, the beam spill comes within a 695ns interval corresponding to single turn extraction 
from the proton accumulator ring.  
This beam spill interval is less than the $\mu^+$ life time
of 2197 ns, but longer than the 26 ns $\pi^+$ life time, allowing good temporal separation of 
the $\nu_\mu$ from $\pi^+$ decay and the $\nu_e \; + \; \overline{\nu}_\mu$ from $\mu^+$ decay.
Third, the recirculating liquid mercury beam stop quickly
absorbs most $\pi^-$ and $\mu^-$ before they decay at rest, 
greatly reduces DIF neutrino production, 
minimizing the associated backgrounds.
The basic parameters of the initial SNS beam dump~\cite{SNSpara} neutrino 
source, and the higher energy source considered at the SNS$^2$ workshop~\cite{SNS2} 
are given in Table~\ref{tab:SNSpars}.
The SNS began construction in 1999, and is scheduled for completion in 2006.  
Full operating intensity should be achieved by 2008, which is an appropriate time scale to
build and commission the detector system described here.
An overview of the SNS site plan is shown in Fig.~\ref{fig:SNSsite}.

Perhaps more important than neutrino oscillation tests will be a wide range of 
neutrino-nucleus cross section measurements that can be made in the neutrino oscillation 
detector discussed here, and in a variety of dedicated single-purpose detectors
placed closer to the SNS beam stop.  
The neutrino energies provided by the beam stop correspond well to the neutrino
flux generated in Supernov\ae, contributing to heavy element synthesis,
the transfer of energy to outer layers of the star, and carrying away
up to 99\%\ of the energy released. 
This paper will also address how other transparent \v{C}erenkov media 
might be used to measure
neutrino-nucleus cross sections in a variety of elements using the SNS neutrino flux.

Previous studies of neutrino physics with the SNS~\cite{Bugg} led to a 
white paper~\cite{WhiteP} 
and a proposal~\cite{D1999} for a large general facility known as ORLaND, the 
Oak Ridge Laboratory for Neutrino Detectors.
A very large \v{C}erenkov detector, ConDOR, of up to 2 kilotons mass was considered
for precision standard model tests using neutrino-electron elastic scattering, 
and to study $\overline{\nu}_\mu \to \overline{\nu}_e$ oscillations to the $10^{-4}$ level.  
The detector presented here is much 
smaller than the CoNDOR/ORLaND design, and is focused on oscillation tests and 
neutrino-nucleus cross section measurements.

The Neutrino Studies at the Spallation Neutron Source (SNS$^2$)
Workshop held August 28-29, 2003 at Oak Ridge~\cite{SNS2}, presents a new opportunity
to bring neutrino physics to the great neutrino source that is being built at the SNS.  
The focus of the workshop was on physics with smaller detectors
within the SNS Target building.  
Small scale segmented and homogenous
detectors for neutrino-nucleus cross section measurements are being considered.

\section{Proposed 250 Ton Detector}

\subsection*{Why 60 meters?}

Discussion has begun on neutrino detectors to be built within the SNS Target Hall itself, 
at about 22 meters from the beam stop~\cite{SNS2}.  
Proposing a detector at 60 or more meters from the beam stop does make sense for several 
reasons. 
The SNS$^2$ site inside the Target Hall near the backward direction would offer about 
4.5$\times$4.5$\times$6.5 m$^3$ for all elements of two
detectors, one homogenous and one fine grained.  
This volume needs to include passive shielding to filter the hadronic component of 
cosmic rays, active shielding to tag muons, and dead space around the fiducial volume.  
Limitations on floor loading plus the passive and active shielding would limit the sensitive
volume of a homogenous detector to at best 5 to 10 m$^3$~\cite{IonSNS2}.
The tight confines would also limit access to detector elements, placing a premium on careful
engineering.

A larger homogenous detector at about 60 meters from the beam stop would be outside the
Target Hall and beyond the extended neutron beam lines in most directions
(see Fig.~\ref{fig:SNSsite}).  
The loss in flux of a factor of about 7.5 would be easily compensated by the ability to make 
a much larger fiducial volume, at least 30 times greater in this proposal.  
Passive shielding, in the form 
of concrete blocks and earth, could be piled as high as needed to reduce cosmogenic backgrounds.
60 meters is, of course, a better distance to test and characterize the 
LSND anti-neutrino oscillation signal.  
The smaller SNS$^2$ homogenous detector within the Target Hall should also be built,
as a near detector for the oscillation test, as a technology testbed for the larger detector, 
and as a less expensive platform for measuring larger neutrino-nucleus cross sections.

\subsection*{The proposed detector}

This paper presents a medium scale detector which would be placed just below the
ground surface at least 60 meters from the neutrino source.  
The detector would not impinge within 10 meters of the Target Hall foundations, 
and requires a final excavation only about 12\%\ of the ORLaND scale.
The Target Hall is shown in relation to the rest of the SNS site in Fig.~\ref{fig:SNSsite}.
The Target Hall is 60 meters wide, with side walls 30 meters on either side of the
beam dump.  
The detector presented here
could definitively test the LSND $\overline{\nu}_\mu \to \overline{\nu}_e$
oscillation result, and also cover a range of interesting 
neutrino-nucleus scattering processes.

Both the original LSND experiment and MiniBooNE have shown that a large imaging
\v{C}erenkov detector can be operated under modest shielding near an accelerator.  
Other successful imaging \v{C}erenkov detectors have operated deep underground
primarily to reduce cosmic ray induced backgrounds to their non-accelerator
physics processes.  

\begin{figure}[htbp]
\includegraphics[width=0.60\textwidth]{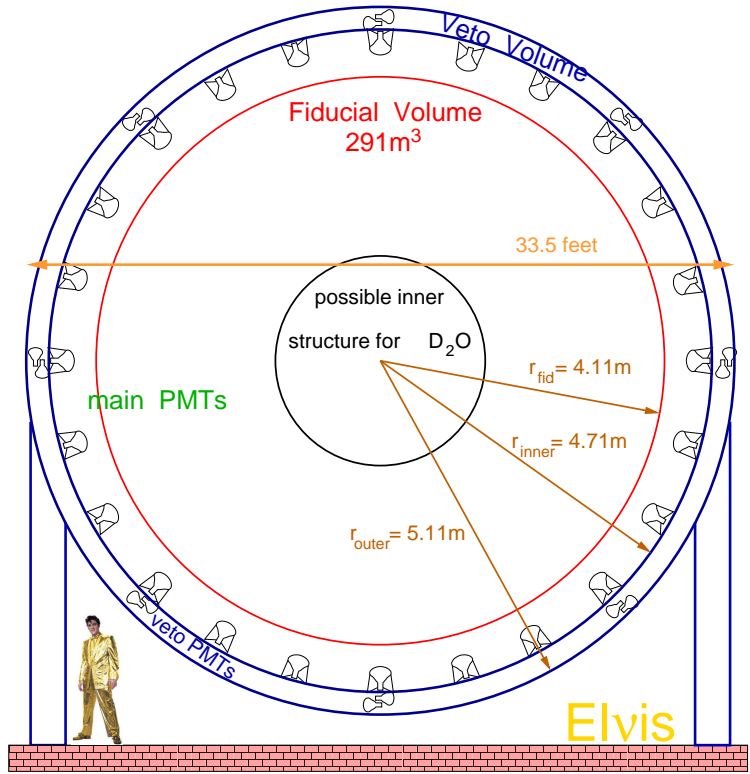}
\caption{\label{fig:Alice} Proposed SNS Detector.
El$\nu$is: (Exchangeable Liquid Neutrino Imaging System)}
\end{figure}

A spherical 250 ton fiducial mass of mineral oil (density 860kg/m$^3$) would have a fiducial 
radius of $4.11{\rm m}$, occupying a volume of 291m$^3$.  LSND and MiniBooNE restrict primary
events to a fiducial volume 35cm inside the PMT faces.
Allowing for an additional 35cm radius for the PMTs and their bases, another 25cm for an active
a veto volume, and 5cm for the combined thickness of the two tank walls, we get an outer 
detector radius of about 5.11 meters.  
The diameter would be about 33.5 feet, compared to the 40 foot diameter of MiniBooNE.
A drawing of the proposed detector is presented in Fig.~\ref{fig:Alice}.

Two newer Hamamatsu photomultiplier tubes~\cite{Hamamatsu} offer better charge and 
time response than the original LSND PMT's (see Table~\ref{tab:PMTs}).  
Both LSND~\cite{LSNDdet} and the Sudbury Neutrino Observatory~\cite{SNOdet}
used R1408 eight-inch PMT's.
The more efficient, better performing R5912 eight-inch PMT has replaced the R1408 in 
Hamamatsu's catalog.  
MiniBooNE uses several hundred R5912 8-inch PMTs among the older LSND tubes.
The newer R8055 has a 13-inch photocathode.  
The number of tubes needed is calculated based on a 25\%\ photocathode coverage 
as used by LSND to detect and resolve 2.2MeV 
gamma rays from neutron capture on free protons.  
MiniBooNE has about 10\%\ photocathode coverage, allowing 
reconstruction of michel electrons from stopping muon decay, but cannot see the
2.2MeV neutron capture signal.  
Table~\ref{tab:PMTs} summarizes the properties of the R5912 and R8055 PMTs, 
and gives the number of each type required for 25\%\ photocathode coverage
in the proposed detector.

A 250 ton (fiducial) detector at the SNS filled with mineral oil, plus butyl-PBD to enhance
scintillation light yield,  would have
$2.15\times 10^{31}$ free protons.
(Which gives about half as many $^{12}$C nuclei, $1.08\times 10^{31}$.  
We also get $8.60\times10^{31}$ atomic electrons.)  We have used a 7\%\ energy 
resolution for electrons at 50 MeV in the calculations below.  
The LSND detector with 1220 
PMTs ultimately achieved an energy resolution of 7\% at the michel endpoint 
of 52.8 MeV~\cite{LSNDlast}.

The original ORLaND facility proposal~\cite{D1999} for the SNS 
called for a hole in the ground 110 feet deep by 78 feet in diameter.  
This size was required to accomodate a 2kton detector plus several smaller detector
for other neutrino-nucleus studies.  
The detector described here requires an excavation of at most 40 feet deep 
by 44 feet diameter, or about 12\% of the volume of the whole ORLaND facility.
 
\subsection*{Shielding issues: passive and active} 

The detector could be built closer to ground level if sufficient earth for a
passive hadron shield is piled overhead.  LSND had a passive shield of 
2000g/cm$^2$~\cite{LSNDdet}, yielding a cosmic muon rate of 4kHz in the detector.  
MiniBooNE has 18 inches of concrete topped by at
3 meters of earth~\cite{MiniBooNE} for a total of at least 4000 g/cm$^2$ of passive shielding.
MiniBooNE experiences a cosmic muon rate of 9.2 kHz.  
The principle deadtime of both LSND and MiniBooNE, which use the same readout electronics,
arises from a 15 $\mu$sec trigger hold-off after each muon.  
Greater passive shielding in the form of additional concrete and earth, plus newer readout 
and greater processing power, should allow a reduction in cosmic muon related deadtime.

LSND had a non-hermetic 15 cm thick liquid scintillator active veto inherited from the 
previous E-645 experiment~\cite{E645}, which has a veto inefficiency for charged particles 
of less than $10^{-5}$.  
Gaps in this active veto led to problematic low energy neutron backgrounds from below which
complicated initial oscillation analysis.  MiniBooNE uses its working fluid of pure mineral oil
without scintillation boosters to form an active veto volume 35 cm thick viewed by 240 PMTs.  
This hermetic system gives a veto efficiency of about 99.97\%, at least an order of magnitude 
poorer than LSND's thinner liquid scintillator veto.

The proposed detector has double one-inch steel walls defining the outside 
of the detector, and the inner active volume.  
The veto and active volumes must be separated so that a highly responsive
liquid scintillator can be used for an efficient veto for any liquid
placed in the inner sensitive volume.
This active veto would be 25 cm thick and hermetic, improving on both LSND and MiniBooNE
experiences.  The veto region can be viewed by several hundred 5-inch PMT's.  
Both the inner and outer wall of the veto region will be
painted reflective white.
The inner surface of the active volume will be painted a non-reflective black.

\subsection*{Detector efficiency $\varepsilon_{\rm rec}$}

The LSND detector, operating with the same event energies expected here, had single electron
reconstruction efficiencies which ranged from 40\%\ to over 50\%\ depending on the specific 
analysis.  The poor duty factor of the LSND meant that stringent cuts needed to be applied
to separate beam related electron events from non-beam backgrounds.  
The much lower duty factor of the SNS neutrino flux should allow a single electron 
efficiency well above 50\%.

MiniBooNE faces a different set of problems while identifying electrons at about 500 MeV.  
Muons at this energy also produce \v{C}erenkov cones, 
and the neutrinos are energetic enough to produce events with $\pi^0$s.
A novel application of neural networks has allowed MiniBooNE to achieve an electron efficiency 
of at least 50\%\ with good muon and $\pi^0$ rejection.

Neutron tagging, required for the anti-neutrino oscillation analysis, was complicated in the
LSND by the backgrounds sneaking in through gaps in veto coverage.  The proposed detector for
the SNS must have an efficient hermetic veto system.

Although electron efficiencies well over 50\%\ should be possible, we have chosen to 
parameterize our ignorance using a reconstruction efficiency $\varepsilon_{\rm rec}$ 
explicitly in all event rates given.

\section{\label{Yuri}SNS Neutrino Fluxes, Shapes and Ratios}
At a nominal beam power of 1.4MW (1300 MeV kinetic energy protons) the SNS
neutron source will also produce $2.92\times 10^{22}$ Decay-At-Rest (DAR)
neutrinos ($\nu_\mu$, $\overline{\nu}_\mu$, and $\nu_e$) per beam year.
The search for \ $\overline{\nu}_\mu \to \overline{\nu}_e$ \ is based on the 
sequence shown here.
\begin{equation*}
\begin{array}{cccc}
{\rm DAR\ in\ beam\ stop}\ \ \  & \ \ \ {\rm oscillation\ in\ flight} \ \ \ & 
\multicolumn{2}{l}{\rm in\ detector} \\
\mu^+ \to {\rm e}^+\; \overline{\nu}_\mu\; \nu_e & 
            \overline{\nu}_\mu \longrightarrow \overline{\nu}_e & 
                  \ \ \ \overline{\nu}_e\; {\rm p} \to {\rm n}\; {\rm e}^+\ \ \  & 
                          \ \ \ {\rm n}\; {\rm p} \to {\rm d} \ \gamma(2.2{\rm MeV}) \\
 & \longleftarrow 60\ {\rm m} \longrightarrow & 
           \multicolumn{2}{r}{{\rm detect} \ e^+ \ {\rm followed\ by} \ \gamma} \\
\end{array}
\end{equation*}

\begin{figure}[htbp]
\includegraphics[width=0.70\textwidth]{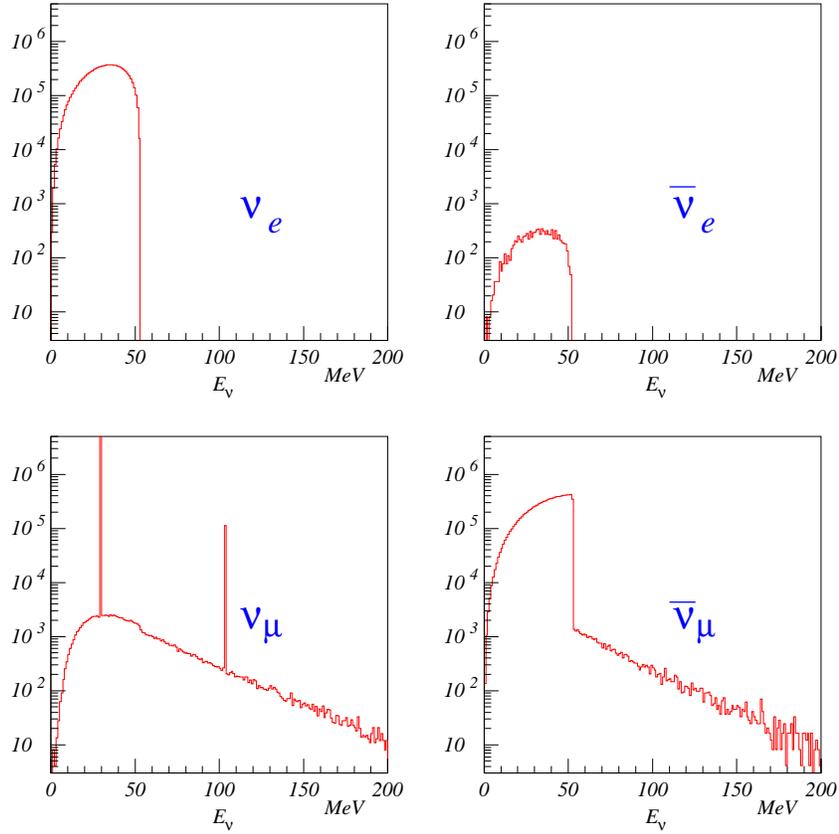}
\caption{\label{fig:yflux} Neutrino energy distributions from the SNS mercury beam stop.
Adapted from Ref.~\cite{YuriNuFact03}.}
\end{figure}

The standard ORLaND neutrino flux plot is shown in Fig.~\ref{fig:yflux}.
Similar plots have
appeared in the ORLaND White Paper~\cite{WhiteP} and various proposal
drafts.  The White Paper~\cite{WhiteP} discusses a design using 1300 MeV
protons on the SNS target, while a 1999 draft for a neutrino
facility~\cite{D1999}
shows this plot, but in the context of a 1000 MeV
proton beam.  
The present SNS Parameters List (May 2003)~\cite{SNSpara} calls for a 1000 MeV beam.  
We use the SNS$^2$ Workshop~\cite{SNS2}
numbers for a 1300 MeV proton beam in what follows. 
However, a 1000 MeV proton beam would give a neutrino
flux only about 5\% lower.  (See Table~\ref{tab:SNSpars}.)

\begin{figure}[htbp]
\includegraphics[width=0.70\textwidth]{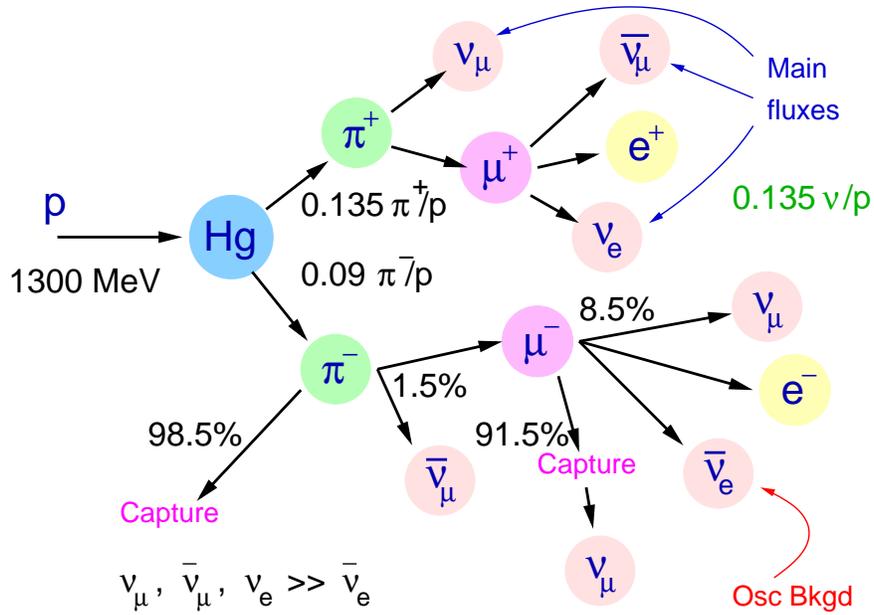}
\caption{\label{fig:SNSstop} Neutrino production in the SNS mercury beam stop.
Adapted from Ref.~\cite{YuriNuFact03}.}
\end{figure}

Each of the four flux components ($\nu_\mu$, $\overline{\nu}_\mu$, $\nu_e$,
$\overline{\nu}_e$) can be analyzed in terms of the parent
processes to see what contribution comes from each decay source.  
The parent processes are sketched  for SNS protons with a 1300 MeV kinetic energy 
in Fig.~\ref{fig:SNSstop}, adapted from Ref.~\cite{YuriNuFact03}.

The total event rate for one year then has essentially three time components.
\begin{itemize}
 \item Beam unrelated (constant) background at a small rate measured in the 
       20$\mu$s before each beam spill
 \item A {\em prompt} component, in time with the beam spill, from
       $\pi^\pm$ DIF events, $\pi^+$ ($\nu_\mu$) DAR events, and the
       majority of the $\mu^-$ DAR (including about 70\% of the
       intrinsic $\overline{\nu}_e$ background.
 \item A {\em decay} component, mostly following the muon DAR time scale,
       including $\nu_e\; ^{12}{\rm C}$ events and the oscillation
       candidates.
\end{itemize}

\begin{figure}[htbp]
\includegraphics[width=0.70\textwidth]{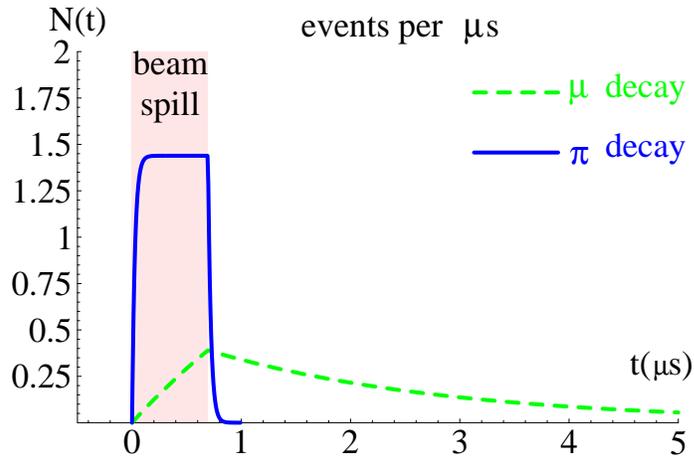}
\caption{\label{fig:osctimes} Neutrino production time distribution.}
\end{figure}

The beam extraction from the storage ring will come in a single turn, 
roughly uniform in intensity over 695 ns~\cite{SNSpara}.  
The fraction of events within a prompt (0 to 695ns) and a muon decay (695ns to 5000ns) 
windows from the components above 
are given in Table~\ref{tab:osctimes}.  
The actual time distributions are illustrated in Fig.~\ref{fig:osctimes}.

We can also estimate the timing of different event types since most of the
$\pi^+$ decay at rest with life time of 26ns.
About 98.5\%\ of the $\pi^-$ capture leaving only 1.5\%\ to decay.  
The $\mu^+$ decay at rest with life time 2197ns, while the $\mu^-$ capturing 
in the beam stop mercury have a life time of 76.2ns~\cite{SMR}.  
A fraction of the $\mu^-$ escape into lighter materials and decay/capture with
the life time closer to the vacuum value.

The intrinsic oscillation background of $\overline{\nu}_e$ from $\mu^-$ DAR
is suppressed by $\pi^-$ and $\mu^-$ absorption in the mercury ($Z=80$)
beam stop.  The total $\overline{\nu}_e$ fraction of $8.5\times 10^{-4}$ 
may be further reduced by requiring that events come after the 695ns beam
spill. 

The neutrino flux simulation shown in Fig.~\ref{fig:yflux} was 
calculated with high-Z reflectors 
around the beam stop.  The final design~\cite{SNSpara} uses beryllium reflectors, 
which will certainly raise the $\overline{\nu}_e$ backgrond, and diminish the potential 
gain from event time analysis.
A new neutrino flux simulation with the final beam stop materials is needed to assess the 
final intrinsic $\overline{\nu}_e$ background, and what might be gained by using 
a time cut on the events to further suppress the $\overline{\nu}_e$ background. 

The number of $\pi^+$ DAR and $\mu^+$ DAR neutrinos produced is
\begin{equation*}
N_\nu = (1.14\times 10^{14}{\rm p/pulse})
(60{\rm pulse/s})(3.16\times 10^7{\rm s/y})(0.135\;\nu/{\rm p}) = 
2.92\times 10^{22}\nu/{\rm beam\ year}
\end{equation*}
At 60m the neutrino flux per beam year would be
$6.45\times 10^{13}\nu/{\rm cm}^2$.

\section{Standard Events in 250 tons of Mineral Oil}

We have used standard techniques to evaluate cross sections and event rates
for neutrino-proton and neutrino-carbon interactions in a mineral oil
filled detector.
Both the $p(\overline{\nu}_e,e^+)n$ and $p(\overline{\nu}_\mu,\mu^+)n$ cross sections are
calculated using the full expressions in Ref.~\cite{vogel}, without low energy approximation.
These calculations are equivalent to the program used by the LSND Collaboration~\cite{louis},
but have been implemented in {\tt Mathematica}, as have most of the
calculations presented in this paper.

The standard electro-weak cross sections~\cite{PDG} have been used to evaluate rates
for neutrino-electron elastic scattering.

The cross section for $^{12}{\rm C}(\nu_e,e^-)^{12}{\rm N}_{\rm gs}$ comes 
from Ref.~\cite{Fuku},
which has been cross checked with the LSND $\nu_e\; C$ results~\cite{LSNDnueC}, and a
table of equivalent cross sections in Ref.~\cite{KLTV}.
The cross section form for $^{12}{\rm C}(\nu_e,e^-)^{12}{\rm N}^*$~\cite{KLTV},
has been cross checked with the LSND $\nu_e\; C$ results~\cite{LSNDnueC}.
The cross sections for $^{12}{\rm C}(\nu,\nu^\prime)^{12}{\rm C}^*_{\rm 15.11}$ and
 $^{12}{\rm C}(\overline{\nu},\overline{\nu}^\prime)^{12}{\rm C}^*_{\rm 15.11}$ are from
Ref.~\cite{Fuku}, and have been compared to measurments by KARMEN~\cite{KARMENnuC}.
The calculation~\cite{Fuku2} of $^{13}{\rm C}(\nu_e,e^-)X$ agrees well with both the
KARMEN result~\cite{KARMENmas} and the LSND final oscillation paper~\cite{LSNDlast}.

Finally, $^{12}{\rm C}(\nu_\mu,\mu^-)X$ and $^{12}{\rm C}(\overline{\nu}_\mu,\mu^+)X$ rates
were scaled from the LSND results~\cite{LSNDnumuC} correcting for the relative DIF fluxes
and fiducial masses.  
Since the SNS DIF neutrino fluxes are also softer, due to the dense high-Z mercury target, 
these last calculations represent only upper limits.
Event rates for fiducial mass of 250 tons of mineral oil,
 are summarized in Table~\ref{tab:concl}.

\section{\label{sec:nuep}The $\overline{\nu}_\mu \to \overline{\nu}_e$ 
Oscillation Signature}

The maximum mixing number of events for the LSND final event sample~\cite{LSNDlast}
was, for $T_e > 20$MeV, 33300 events.
The maximum mixing events for a 250 ton (fiducial) detector at 60m from
the SNS beam stop, using LSND reconstruction efficiency of 42\%, would be
49500 events in one beam year.  We have used the cross section for 
$p(\overline{\nu}_e,e^+)n$ from Refs.~\cite{vogel,louis}.
Table~\ref{tab:LSNDvSNS} gives a comparison of the LSND parameters with
the detector proposed here.

\begin{figure}[htbp]
\includegraphics[width=0.50\textwidth]{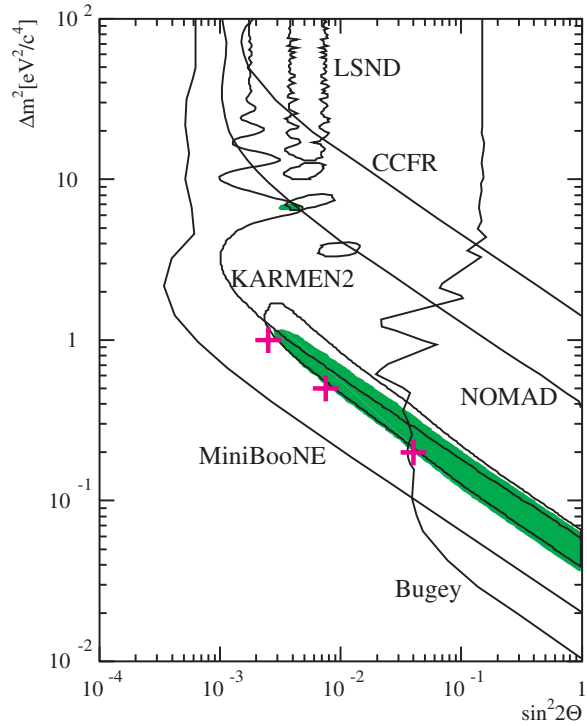}
\caption{\label{fig:church}Region in $\Delta m^2$ vs.~$sin^2 2\theta$ from a combined
analysis of LSND and KARMEN~\protect\cite{church}.  
The crosses show three conservative oscillation
parameter sets considered in this paper.}
\end{figure}

To calculate specific oscillation event rates, a set of {\em very conservative}
oscillation parameters are taken from the combined analysis of both LSND and KARMEN
by Church {\em et al.}~\cite{church}.
The points in $\sin^2 2\theta$ and $\Delta m^2$ used in the calculations here
are shown as crosses ($+$) in Figure~\ref{fig:church}.  
These points are, as can clearly be seen, conservative.

\begin{figure}[htbp]

\includegraphics[width=0.45\textwidth]{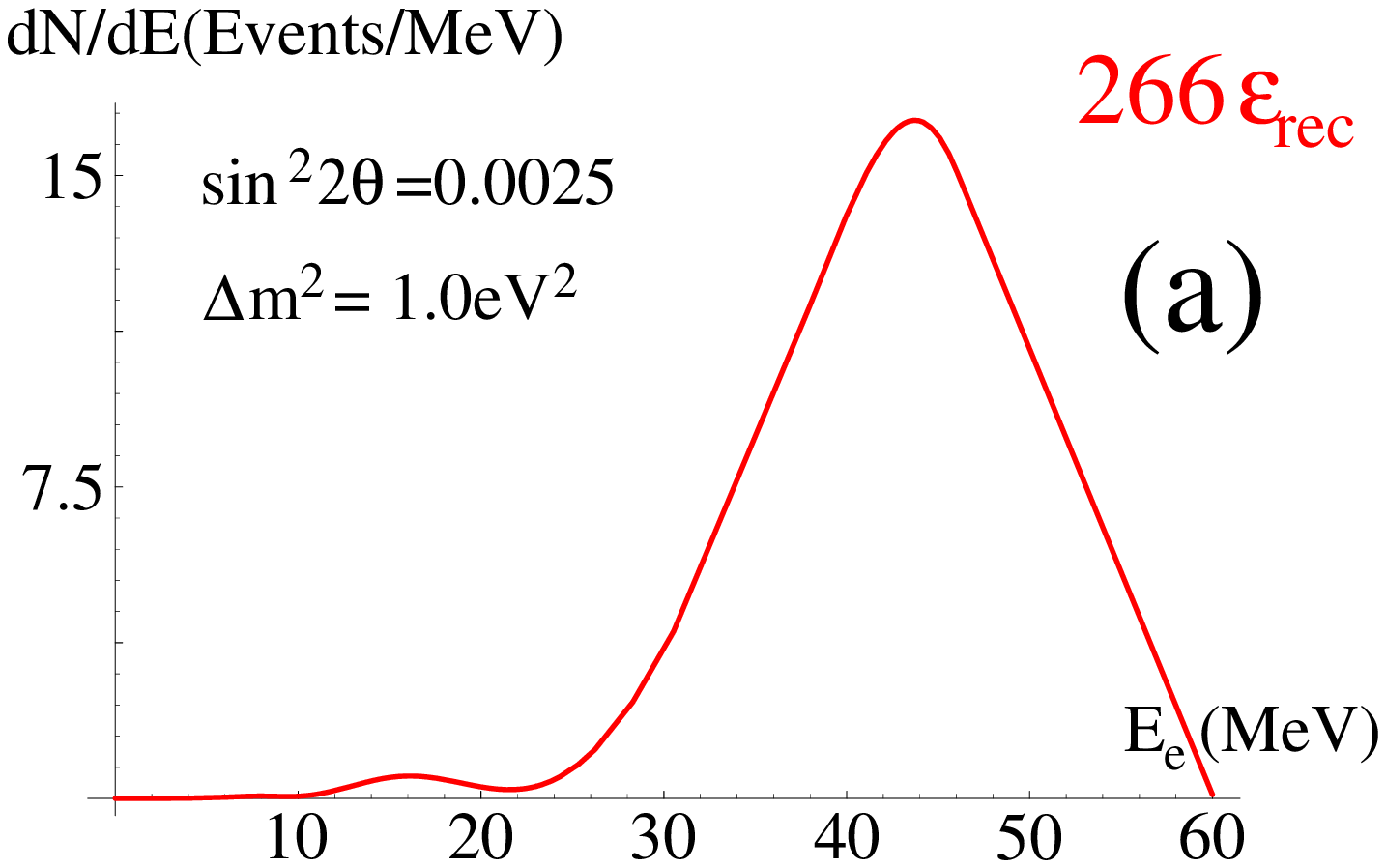}
 
\vskip 0.1in
\includegraphics[width=0.45\textwidth]{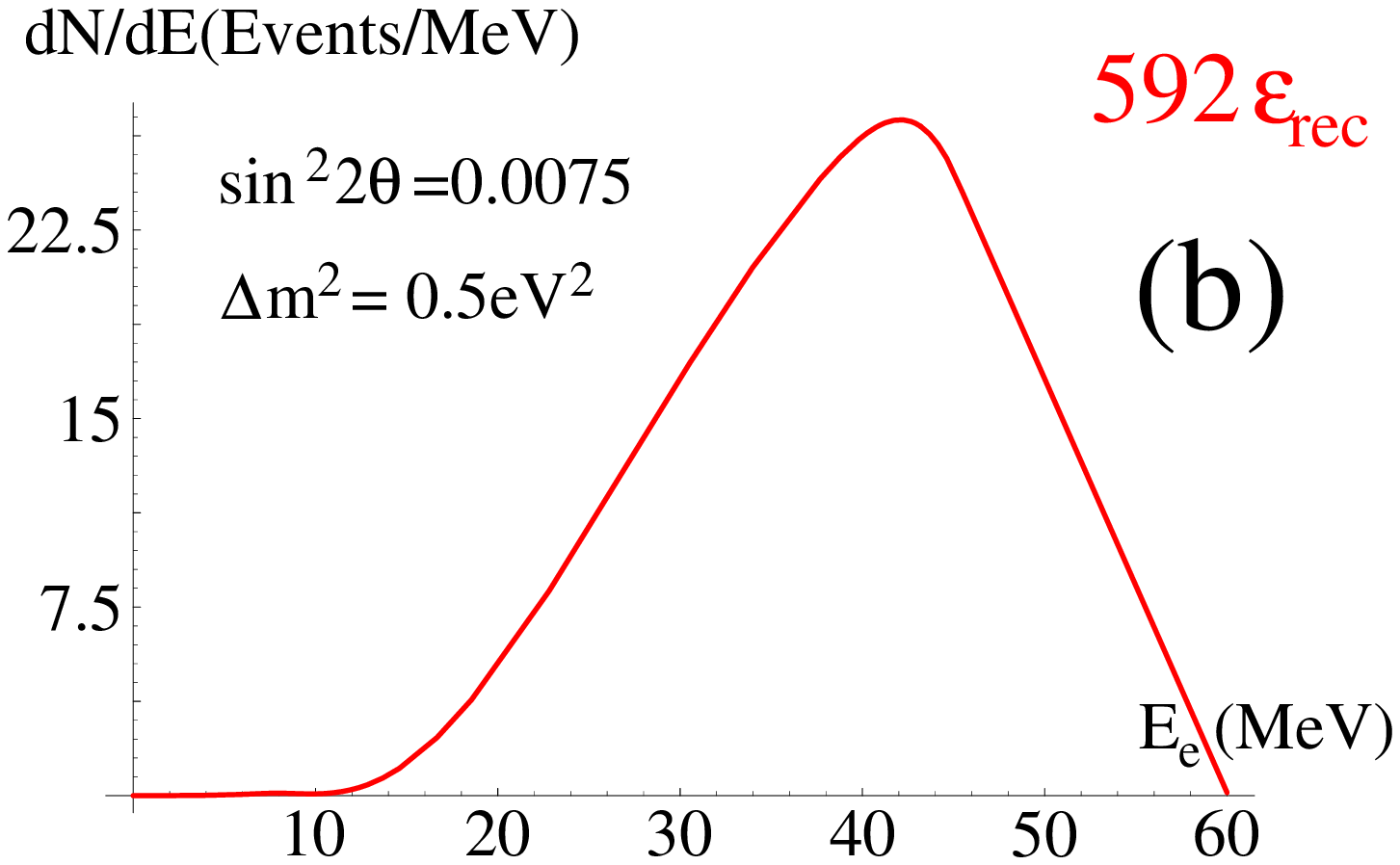} \ \ \ 
          \includegraphics[width=0.45\textwidth]{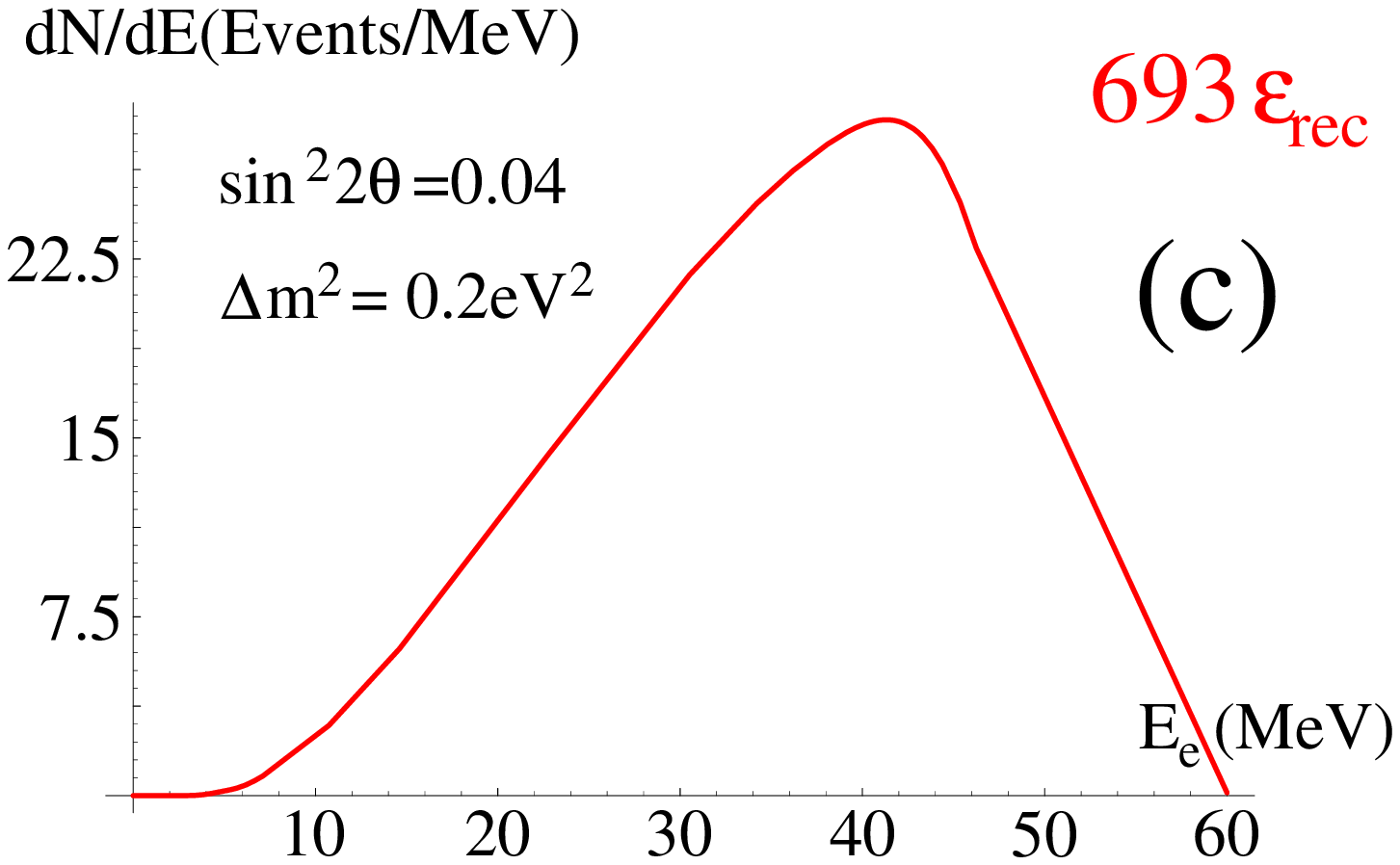}
\caption{\label{fig:3plots} DAR $\overline{\nu}_\mu \to \overline{\nu}_e$ 
Oscillation Event Energies in 250 ton SNS Detector
with Energy Smearing (7\% at 50MeV) }
\end{figure}

The three plots in Figure~\ref{fig:3plots} show the smeared (7\%\ at 50 MeV) energy 
distribution of events for each of the 3 cases considered.  
These are graphed directly in terms of events per MeV of energy per beam year.

\section{Oscillation Backgrounds}

\begin{figure}[htbp]
\includegraphics[width=0.65\textwidth]{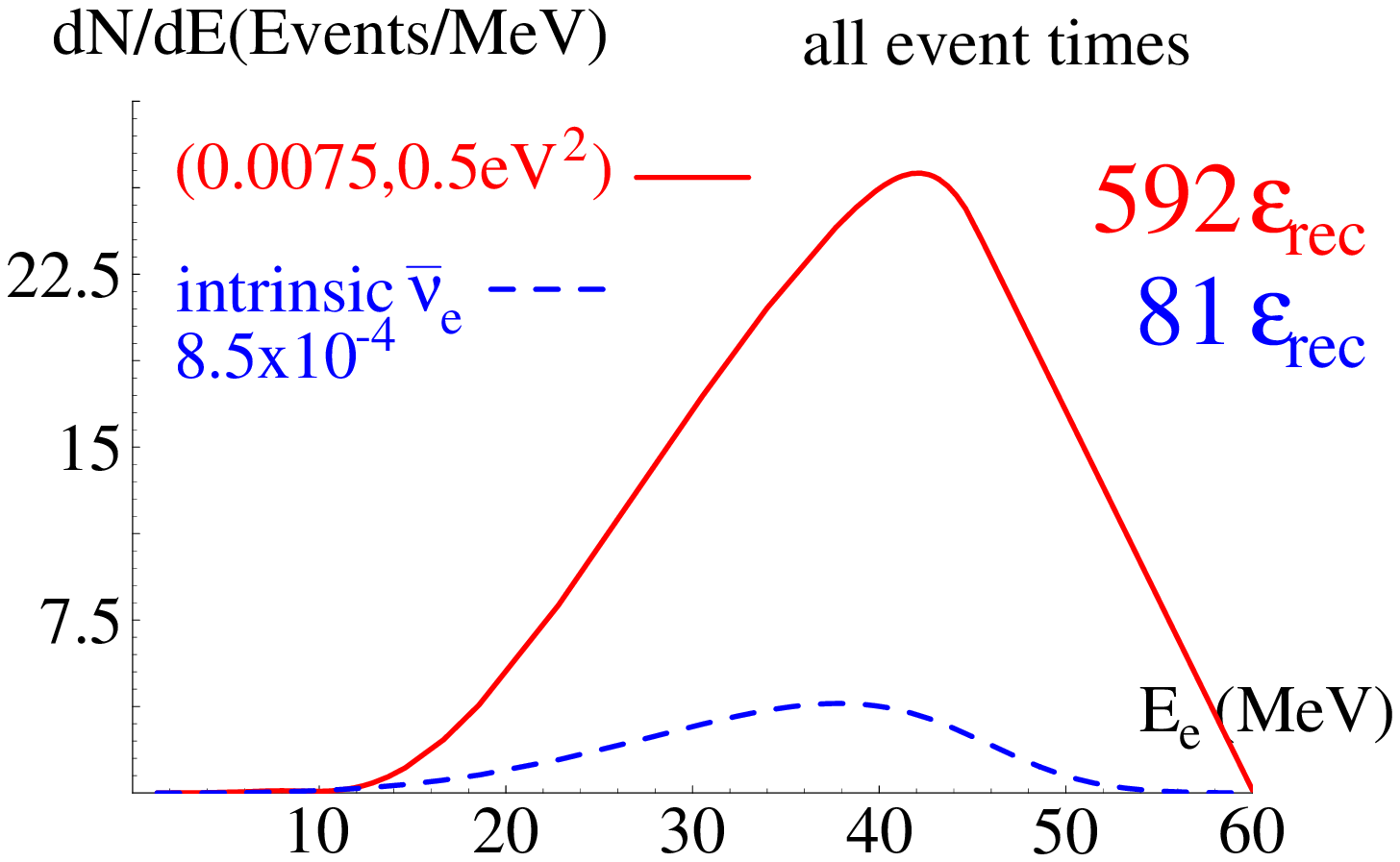}

\caption{\label{fig:nuebar} Intrinsic  $\overline{\nu}_e$ Component and
           $\overline{\nu}_\mu\to\overline{\nu}_e$ Oscillations with
           $\sin^2 2\theta = 0.0075$, $\Delta m^2 = 0.5{\rm eV}^2$.}
\end{figure}

The largest background comes from the residual $\overline{\nu}_e$ component of the
beam.  Fig.~\ref{fig:nuebar} shows the signal for the parameters of Fig.~\ref{fig:3plots}(b)
with the intrinsic $\overline{\nu}_e$ background.  

The other SNS beam related backgrounds come from misidentified DIF events, where the
muon was missed.  The energy distribution of these events comes simply from
the michel electron spectrum.  We can calculate the $p(\overline{\nu}_\mu,\mu^+)n$
background directly by requiring a muon kinetic energy below 3 MeV.  The backgrounds
from $\nu_\mu\;^{12}{\rm C}$ and $\overline{\nu}_\mu\;^{12}{\rm C}$ are estimated
from the scaled LSND backgrounds~\cite{LSNDnumuC}.  
The total of these DIF backgrounds is less than 
$5\;\varepsilon_{\rm SNS}$ events per beam year.

The LSND experiment ran for approximately 15 months total over the years
1993-98, with a duty factor of about 0.06 imposed by the long beam pulse
characteristic of LAMPF.  
The SNS beam duty factor is $4.2\times 10^{-5}$, but the experiment must wait 
about 5 microseconds after each pulse starts to get most of the neutrino events
from $\mu^+$ DAR.  This gives an {\em experimental} duty factor more like
$3\times 10^{-4}$.
We can make a rough estimate the SNS beam-off background by scaling the 
LSND beam off background.  The scaling factor is
{\footnotesize
\begin{equation*}
{\frac{ \left[{\rm duty\ factor(SNS)}\right]\;
         \left[{\rm live\ time(SNS)}\right]\;M_{\rm SNS}\;
         \varepsilon_{\rm SNS}}{
 \left[{\rm duty\ factor(LSND)}\right]\;
\left[{\rm live\ time(LSND)}\right]\;M_{\rm LSND}}\;\varepsilon_{\rm LSND}} = 
{\frac{(3\times 10^{-4})(12{\rm mo.})(250{\rm tons})\;\varepsilon_{\rm SNS}}{
(0.06)(15{\rm mo.})(86.5{\rm tons})(0.42)}} = 0.0275\;\varepsilon_{\rm SNS}
\end{equation*}
\small}

\noindent
The LSND experiment had 107 beam off events in the $R_\gamma>1$ sample, 
which would
give just $3\;\varepsilon_{\rm SNS}$ background beam off 
events in the SNS detector for one beam year.

The $\overline{\nu}_\mu \to \overline{\nu}_e$ oscillation backgrounds and signals
are summarized in Table~\ref{tab:concl2}.

\section{$\nu_\mu\to \nu_e$ Oscillations}

The LSND experiment was able to use the 3\% $\pi^+$ DIF flux of $\nu_\mu$'s
to also search for $\nu_\mu\to \nu_e$ oscillations.  
The relatively high DIF flux was achieved by 
having a low-$Z$ moderator (water) before a decay path and the final beam
stop in 1993-95.  The LSND {\em neutrino} oscillation paper~\cite{LSNDion}
reported a result of modest statistical significance, oscillation 
probability of $(2.6\pm 1.0\pm 0.5)\times 10^{-3}$, even with their higher
DIF $\nu_\mu$ flux.  
With the lower DIF flux fractions at the SNS, the DIF oscillation rate is 
estimated at 6$\varepsilon_{\rm rec}$ events with a negligible background.

It has been suggested~\cite{D1999} that DAR 29.8MeV $\nu_\mu$'s from
$\pi^+\to \mu^+\; \nu_\mu$ DAR could be detected after oscillation by
the process
{\footnotesize
\begin{equation*}
\begin{array}{cccc}
{\rm DAR\ in\ beam\ stop}\quad  & \quad {\rm oscillation\ in\ flight} \quad & 
\multicolumn{2}{l}{\rm in\ detector} \\
\pi^+ \to \mu^+\; \nu_\mu\ & 
            \nu_\mu \longrightarrow \nu_e & 
                 \ \ \nu_e\; ^{12}{\rm C} \to e^-\; ^{12}{\rm N}_{\rm gs} \ \ & 
         \ \ \;^{12}{\rm N}_{\rm gs}\; \to \; ^{12}{\rm C} \; e^+ \; \nu_e\\
 & \longleftarrow 60\ {\rm m} \longrightarrow & 
             \multicolumn{2}{r}{{\rm detect}\ e^+\ {\rm within\ 50ms}} \\
\end{array}
\end{equation*}
}

The oscillation event electrons come at about (29.8 - 16.8)MeV = 13MeV, 
spread by the energy resolution.  
So we are back in the traditional
particle physics situation of looking for a ``bump'' on a known spectrum.
Using the analytic cross section~\cite{Fuku} 
for $^{12}{\rm C}(\nu_e,e^-) ^{12}{\rm N}_{\rm gs}$ we can calculate 
one of the backgrounds and the signal.

The largest beam related background is from DAR $\nu_e$ in
the beam coming from $\mu^+\to e^+\; \nu_e\; \overline{\nu}_\mu$ on a
muon DAR time scale.  Only about 14.3\%\ of these events will come
during the actual beam spill where we expect the signal from the oscillation
of monochromatic 29.8MeV $\nu_\mu$ from $\pi^+$ DAR.

\begin{figure}[htbp]
\includegraphics[width=0.45\textwidth]{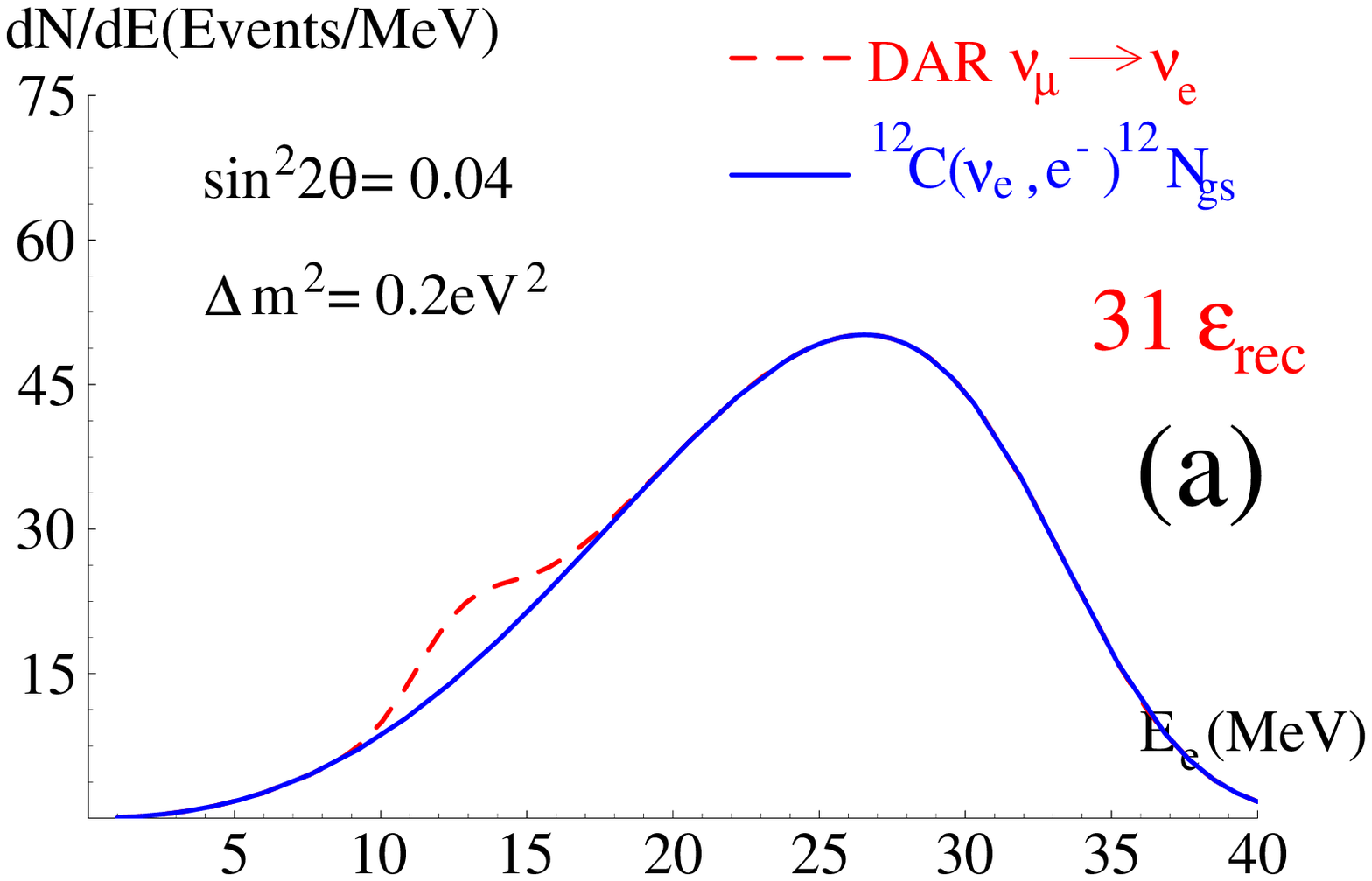}\ \ \ \ 
           \includegraphics[width=0.45\textwidth]{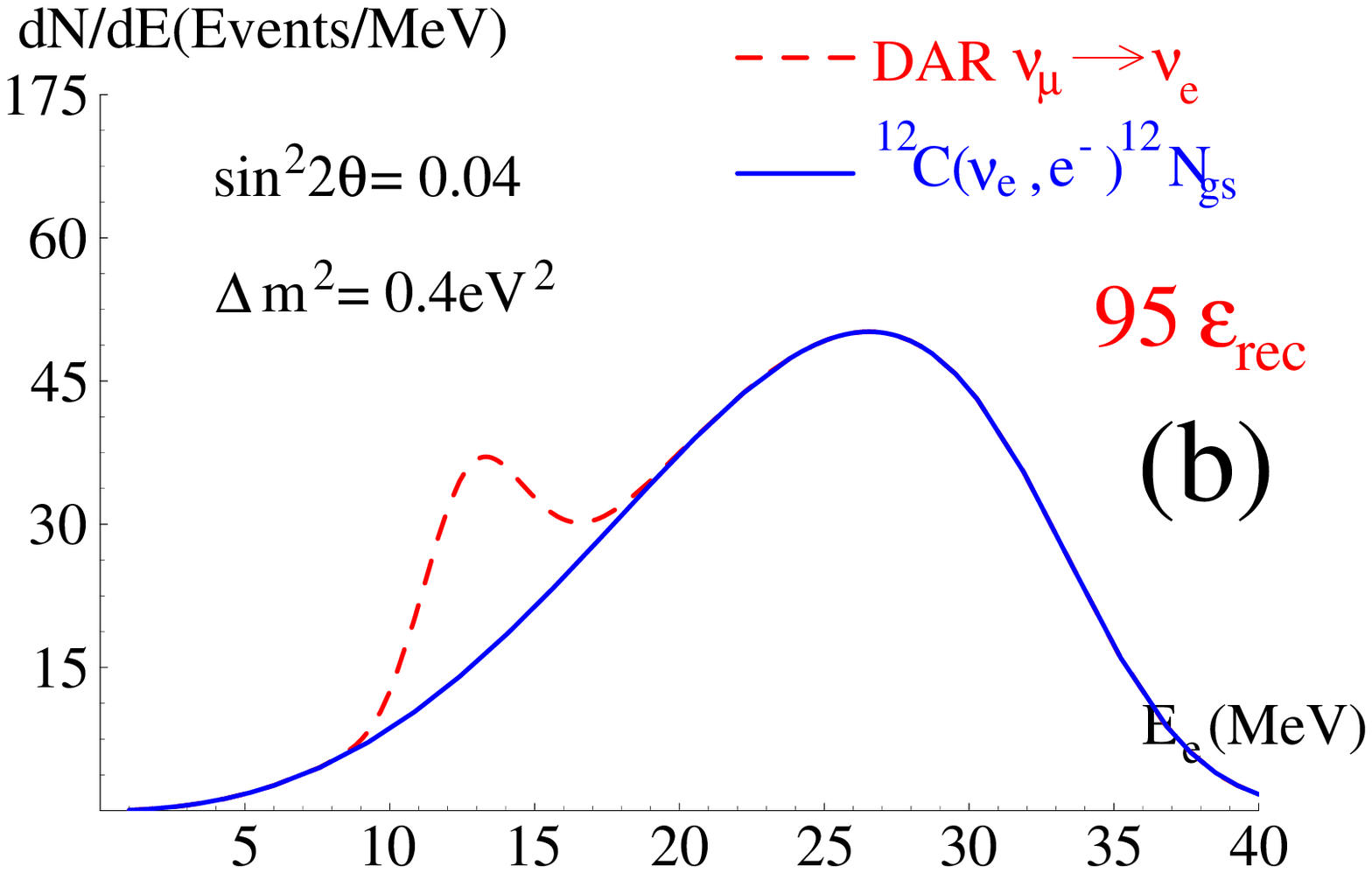}
\caption{\label{fig:nsosc} DAR $\nu_\mu\to \nu_e$ Oscillation Signals for two possible
sets of oscillation parameters.  (a) For $\sin^2 2\theta = 0.04$, $\Delta m^2 = 0.2{\rm eV}^2$,
(b) for $\sin^2 2\theta = 0.04$, $\Delta m^2 = 0.4{\rm eV}^2$.  The background from
$^{12}{\rm C}(\nu_e,e^-)^{12}{\rm N}_{\rm gs}$ is shown as the solid blue curve.
The backgrounds from $^{12}{\rm C}(\nu,\nu^\prime)^{12}{\rm C}^*_{\rm 15.11}$ of 
$\sim 15$ events peaked at 15.11 MeV is not shown.
}
\end{figure}

We have plotted the signal and the $^{12}{\rm C}(\nu_e,e^-) ^{12}{\rm N}_{\rm gs}$
background in Fig.~\ref{fig:nsosc} for two different $\Delta m^2$ at 
$\sin^2 2\theta = 0.04$ which give a possible signal.  
Over the combined LSND/KARMEN
allowed region~\cite{church} the size of the bump varies from
a few up to $100\varepsilon_{\rm rec}$ events.

We have also calculated~\cite{Fuku} the cross sections 
for $^{12}{\rm C}(\nu,\nu^\prime)^{12}{\rm C}^*_{15.11}$ where the 15 MeV
gamma-ray could potentially present a large background to 13 MeV electron in this process.
The KARMEN Collaboration measured~\cite{KARMENmas,KARMENnuC} the
sum of the
$\overline{\nu}_\mu$ and $\nu_e$ induced NC transitions to be
$(10.9\pm 0.7\pm 0.8)\times 10^{-42}{\rm cm}^2$, consistent with our
calculation which gives $9.86\times 10^{-42}{\rm cm}^2$.

The total $^{12}{\rm C}(\nu,\nu^\prime)^{12}{\rm C}^*_{15.11}$ background
in the beam window would be 14.3\%\ of the $\mu^+$ DAR events plus 96.3\%\ 
of the $\pi^+$ DAR events, for a total of about
$2760\; \varepsilon_{\rm rec}$ events.
The LSND \ $^{12}{\rm C}(\nu_e,e^-) ^{12}{\rm N}_{\rm gs}$ \ analysis 
associated the $^{12}{\rm N}_{\rm gs}$ decay betas with an efficiency of 
about 60\% and an accidental background rate of about 0.5\% of the remaining 
events.  If we accept this accidental background rate, we would have about 
$15\; \varepsilon_{\rm rec}$ background events from accidentally associated 
$^{12}{\rm C}(\nu,\nu^\prime)^{12}{\rm C}^*_{15.11}$ background.

The DAR $\nu_\mu \to \nu_e$ process gives a 
signal bump of $(2 \to 100)\; \varepsilon_{\rm rec}$ 
events with a beam related background bump of $15 \; \varepsilon_{\rm rec}$
events from $^{12}{\rm C}(\nu,\nu^\prime)^{12}{\rm C}^*_{15.11}$ on top
of a smooth background from $^{12}{\rm C}(\nu_e, e^-) ^{12}{\rm N}_{\rm gs}$
of about $15 \; \varepsilon_{\rm rec}$ events per MeV.  This DAR
$\nu_\mu\to\nu_e$ measurement just might be 
possible if the mixing parameters are large enough.

\section{Other Liquids for the Detector}
There are other amusing materials that one could use to fill all, or only a portion, of
a large neutrino detector at the SNS.  
Then we could measure astrophysically interesting neutrino interactions 
with D, C, N, O, Cl, Br, I,
etc.~and others within the ``imaging \v{C}erenkov'' detector technology.

With some imagination, and some materials testing, we could measure
a variety of neutrino-nucleus cross sections with the SNS DAR neutrino fluxes.
So long as the material is relatively inexpensive, respects the tank materials,
and is not too hazardous, the concept of imaging \v{C}erenkov can be taken well
beyond the choice between water and oil.  Table~\ref{tab:other} gives a short list of 
other potential liquids.  Certainly there are many other liquids to consider as well.

Mineral oil (LSND and MiniBooNE) and water (SuperK, SNO, IMB) have relatively well
known properties as a \v{C}erenkov medium and for the liquid handling and purification
requirements.  
SNO is also gaining experience with the use of dissolved salts (NaCl) to tag the 
presence of low energy neutrinos.
Chlorine ($^{35}$Cl) has a high absorption cross-section for thermal neutrons, giving
a gamma ray cascade peaked at around 8 MeV~\cite{SNOdet}.
An SNS detector could use low concentrations of salts (as in SNO) to enhance neutron
tagging in water, or could use higher concentrations of salt to provide other nuclear targets.

The simplest compound with nitrogen, ammonia (${\rm N H_3}$), has a boiling point of
about $-33^\circ$C.  
Mixtures of ammonia and water can give boiling points above 0$^\circ$C.
For example, a mixture of 90g ${\rm N H_3}$ per 100g ${\rm H_2 O}$ boils at 10$^\circ$C.  
The DAR flux averaged cross section for $^{14}{\rm N}(\nu_e, e^-)^{14}{\rm O}$ is 
calculated at $29\times 10^{-42}{\rm cm}^2$~\cite{AandB},
about twice the total $\nu_e\;^{12}$C cross section, and about three times the
calculated total $\nu_e\;^{16}$O cross section~\cite{AandB}.  
Natural nitrogen is almost all $^{14}$N with only 0.37\%\ of $^{15}$N.

Table~\ref{tab:other} gives other organic compounds containing at most one additional
element.  These liquids are transparent, have high index of refraction.  
Basic properties were gleaned from the CRC Handbook of Chemistry and Physics~\cite{CRC},
the International Labor Organization International Safety Cards~\cite{ILO}, 
the Atomic and Nuclear Properties of Matter from the PDG~\cite{PDGanpm}, and the NIST
Stopping Power and Range Tables for Electrons~\cite{ESTAR}.
Several of these materials are available in industrial amounts.
Order by the ``rail car'' unit by checking the web for suppliers and quotes.

Tetrachlorethene (${\rm C_2 Cl_4}$) was used in the Homestake solar neutrino detector
by Ray Davis~\cite{RDavis}.  
The common industrial use of tetrachlorethene was as 
a dry cleaning fluid, so it respects most artificial fibers used in clothing.  
The vapor of tetrachlorethene is heavier than air, and it suspected to be a carcinogen.  
There is no tolerance for any amount of spillage.  
Natural chlorine is roughly $3/4$ $^{35}{\rm Cl}$ 
and $1/4$ $^{37}{\rm Cl}$.  As noted above, $^{35}$Cl has a high absorption cross-section 
for thermal neutrons, giving a gamma ray cascade peaked at around 8 MeV.
The DAR flux averaged cross section for $^{37}{\rm Cl}(\nu_e, e^-)^{37}{\rm Ar}$ is 
calculated at $180\times 10^{-42}{\rm cm}^2$~\cite{KandO}.  
We would also need calculations of the $\nu_e\;^{35}$Cl cross section.

Methylene bromide (${\rm C H_2 Br_2}$) is a commonly used industrial degreaser 
available in large amounts.  
Ethylene dibromide (${\rm C_2 H_4 Br_2}$, aka EDB) is used as an ``octane'' booster in 
aviation gas, and as a degreaser.  
Tribromoethene (${\rm C_2 H Br_3}$) is rare and toxic.  
Bromine naturally occurs as about equal portions of $^{79}$Br and
$^{81}$Br. 
Ref.~\cite{solarBr} proposed using any of these same three liquids in the
380 m$^3$ tank of the Homestake Chlorine experiment~\cite{RDavis} to also measure solar 
neutrinos.
The $^{81}$Kr would have been detected by resonance ionization spectrometry rather than by
radiochemical means.
The DAR flux averaged cross section for $^{81}{\rm Br}(\nu_e, e^-)^{81}{\rm Kr}$ is 
calculated at $450\times 10^{-42}{\rm cm}^2$~\cite{KandO}.  Cross section estimates are
also needed for $\nu_e\;^{79}{\rm Br}$.

Ethyl iodide (${\rm C_2 H_5 I}$) is used in small amounts with $^{14}$C as a radiochemical tag
in metabolism studies.  Methylene iodide (${\rm C H_2 I_2}$) is prized for its high index of
refraction, used in the gemstone industry for immersion tests to quickly sort materials.
Both materials will darken (iodine liberated) upon exposure to light.  Natural iodine is 100\%
$^{127}$I.  Cross sections for $\nu_e\;^{127}{\rm I}$ will be discussed in the next section
where event rates in a methylene iodide detector are presented.

All organic compounds listed are solvents which could easily be doped with butyl-PBD to enhance
scintillation light yield.
C$_2$Cl$_4$ could be added to provide $^{35}$Cl to tag neutrons associated with primary
neutrino interactions.

Finally, heavy water D$_2$O could be placed in a transparent subvolume within a water filled
detector.   D$_2$O does have a specific gravity 10.5\%\ greater than water, so this would 
need to be a substantial transparent containment vessel.  
This same system, if strong enough could also hold subvolumes of other expensive liquids 
where the cross sections are expected to be high enough to eliminate the need for filling the
full fiducial volume.
 
How would Hamamatsu feel about having their PMT's in ammonia at -40$^\circ$C?
Have people tested tubes in any other materials like these?  Cables?  Paints?
Other tank materials?  A good summer student project would be to measure
optical properties and PMT responses for these and other organic solvents.

\section{Rates for Methylene Iodide ${\rm C H_2 I_2}$}

The charged current cross section for \ $\nu_e\; ^{127}{\rm I}$ \ could be measured quickly 
by filling the inner volume of the detector with methylene iodide.  Within the fiducial volume
there would be 967,000 kg or 967 tons of material.  
(Our ``250 ton'' size only applies to mineral oil.)
The fiducial volume then contains $2.17\times 10^{30}$ CH$_2$I$_2$ molecules.
With 2 iodine atoms per molecule, there are \ $4.35\times 10^{30}$ iodine atoms 
in the fiducial volume.

Iodine has just one stable isotope, namely $^{127}{\rm I}$.  The threshold neutrino energy for
$^{127}{\rm I}(\nu_e,e^-)^{127}{\rm Xe}$ is 0.789MeV.  
For excitation energies above
about 8 MeV the $^{127}{\rm Xe}$ nucleus is unstable to neutron emission.
An experiment run at LANL~\cite{I127exp} measured the rate for 
\begin{equation*}
\nu_e\; ^{127}{\rm I} \to \;^{127}{\rm Xe}_{\rm bound\ states} e^- \qquad \Longrightarrow 
\qquad ^{127}{\rm Xe}_{\rm gs}
\end{equation*}
where the $^{127}{\rm Xe}_{\rm gs}$ was detected radio-chemically by its 36.4 day half-life 
electron capture decay back to $^{127}{\rm I}$.  The measured cross section is
$(2.75\pm 0.84 \pm 0.24) \times 10^{-40}{\rm cm}^2$.
A theoretical calculation~\cite{I127theo} for the same exclusive process 
gave $(2.1 \to 3.1) \times 10^{-40}{\rm cm}^2$.
Ref.~\cite{KandO} calculates that the {\em total} $\nu_e\;^{127}$I cross section is about 
70\% higher than that for the bound states.
The proposed SNS detector discussed here could get 131,000$\;\varepsilon_{\rm rec}$ 
$^{127}{\rm I}(\nu_e, e^-)X$ events per beam year.

The event rate for $\nu_e\;{\rm C}$ charged currents can be calculated, 
being careful that there is but one carbon atom for every two iodine atoms in
methylene iodide.
The event rates for the methylene iodide filled detector are summarized in Table~\ref{tab:nuI}.
The 131,000 $\nu_e\;^{127}{\rm I}$ events should stand out clearly.

There do not appear to be any short lived betas in the $A = 127$ system (half-life 
less than 100 milliseconds), but the detector could tag additional excited states through 
prompt neutron emission or gammas.  Radiochemical experiments can measure the  
cross section which leads to the $^{127}{\rm Xe}$ bound states, but an imaging \v{C}erenkov
detector can measure differential cross sections in electron energy and angle, as well as tag 
excited states with emitted neutrons and/or gammas.
The SNS neutrino physicists need guidance on both semi-inclusive and total cross sections
for $\nu\;^{127}{\rm I}$ charged and neutral currents.  
The detector could be sensitive to prompt neutrons and gammas, 
as well as delayed gammas and betas on the time scales from 10 microseconds to 
100 milliseconds.

Although the event rate for methylene iodide is quite large, other materials suggested in
the previous section would give between a few thousand and tens of thousands of events
per beam year.

\section{Conclusion}

$\overline{\nu}_\mu\to\overline{\nu}_e$ 
oscillation event rate for $\Delta m^2 < 0.5{\rm eV}^2$ will be a factor 
of about {\bf four} enhanced relative to conventional processes due to
the doubling of $L$ from 30 meters used by LSND to 60 meters as proposed here.

A 250 ton detector placed 60 meters from the SNS mercury beam stop can easily
confirm or refute the LSND $\overline{\nu}_\mu\to\overline{\nu}_e$ oscillation
signal within one nominal beam year of data taking.  The backgrounds are much less
than at the LSND experiment, and signals over the LSND/KARMEN low $\Delta m^2$
allowed region will be several hundred events.  The same experiment can 
improve previous measurements of $\nu_e\;^{12}{\rm C}$ charged current, 
$\nu\; ^{12}{\rm C}$ neutral current, and $\nu\; e$ elastic cross sections
with DAR neutrinos.  Moving the detector futher away from the beam stop, out to 
100 meters, decreases the rate from conventional processes as $1/r^2$, while oscillation
signal rates remain roughly constant for $\Delta m^2 \le 0.3{\rm eV}^2$.

The detector can also study other neutrino-nucleus 
cross sections with high statistics using other liquids.  
This can be a long lived experiment as different liquids are studied in turn.

\section*{Acknowledgements}

First, I would like to thank Prof.~Darrel Smith and my colleagues at Embry-Riddle Aeronautical 
University for providing me a new home in retirement.  Thanks also to the organizers of the
Neutrino Studies at the Spallation Neutron Source Workshop held August 28-29, 2003 
at Oak Ridge National Lab for giving me a chance to present these ideas.  Their work
to prepare an SNS$^2$ proposal will bring neutrino physics to what soon will be our best
neutrino source.  Also, my thanks to Yuri Efremenko, Bill Louis, and Ion Stancu for their
comments and encouragement.  Finally I acknowledge all the work which has gone before 
on neutrino physics at the SNS, particularly for the persistence and leadership of
Frank Avignone.

\newpage

%
%
\begin{table}[htbp]
\caption{\label{tab:SNSpars} Key SNS Parameters.}

\vskip 0.2in
\begin{tabular}{|c|c|c|c|}
\hline
{\bf SNS Parameter}        & \ \ \ {\bf SNS Parameters List~\protect\cite{SNSpara}} \ \ \ & 
  \ \ \ {\bf SNS$^2$ Workshop~\protect\cite{SNS2}} \ \ \ \\
\hline
beam power on target       & 1.4 MW & 1.4 MW \\
beam energy on target      & 1 GeV & 1.3 GeV \\
repetition rate            & 60 Hz & 60 Hz \\
protons per pulse          & $1.6\times 10^{14}$ & $1.14\times 10^{14}$ \\
extracted pulse length     & 695 nsec & 695 nsec \\
target                     & mercury & mercury \\
beam spot on target        & 7cm $\times$ 20cm (V$\times$H) & 7cm $\times$ 20cm (V$\times$H) \\
\ \ DAR neutrinos per proton \ \  & 0.098$\nu$/p & 0.135$\nu$/p \\
DAR neutrinos per beam year     & $2.78\times 10^{22}\;\nu$ & $2.92\times 10^{22}\;\nu$ \\
\hline
\end{tabular}
\end{table}

%
%
\begin{table}[htbp]
\caption{\label{tab:PMTs} Two Hamamatsu Phototubes for the SNS \v{C}erenkov Detector}

\vskip 0.2in
\begin{tabular}{|c|cc|}
\hline
      & \ \ \ \ \ \ {\bf R5912} \ \ \ \ \ \  & \ \ \ \ \ \  {\bf R8055} \ \ \ \ \ \  \\
\hline
height                                 & 290 mm & 332 mm \\
\hline
diameter                               & 202 mm & 332 mm \\
\hline
effective diameter                     & 190 mm & 325 mm \\
\hline
quantum efficiency at 390nm            & 22\%   & 20\% \\
\hline
spectral response                      & \multicolumn{2}{|c|}{300 to 650 nm} \\
\hline
peak sensitivity                       & \multicolumn{2}{|c|}{420 nm} \\
\hline
typical HV for $10^7$ gain             & \multicolumn{2}{|c|}{1500 volts} \\
\hline
transit time spread (FWHM)             &  2.4 ns   & 2.8 ns \\
\hline
number for 25\%\ photocathode coverage & 2106 & 720 \\
\hline
\end{tabular}
\end{table}

%
%
\begin{table}[htbp]
\caption{\label{tab:osctimes} Event fractions in time windows}

\vskip 0.2in
\begin{tabular}{|l|c|c|c|}
\hline
\ \ {\bf event types}                 & \ \ \ \ $0\to 0.695\mu$sec \ \ \ \ &
      \ \ \ \  $0.695\to 5\mu$sec \ \ \ \ &  \ \ \ \  $> 5\mu$sec \ \ \ \  \\
\hline 
\ \ $\pi^+$ DAR and $\pi^\pm$ DIF events          & 96.3\% & 3.7\% & 0 \\ 
\hline
\ \ $\mu^+$ DAR  and oscillation candidates & 14.3\% & 73.6\% & 12.1\% \\
\hline
\end{tabular}\end{table}

%
%
\begin{table}[htbp]
\caption{\label{tab:concl} Event Rates in 250tons of Mineral Oil}

\vskip 0.2in
{\footnotesize
\begin{tabular}{|llrr|}
\hline
               & \ \ Neutrino \ \ & \ \ $\nu$ Source \ \ & \ \ Total Events \ \ \\
\ \ Process \ \ & \ \ Source \ \   & \ \ Life Time  \ \    & \ \ per Beam Year \ \ \\
\hline
$\nu_\mu\; e$      & \ \ $\pi^+$ DAR & 26ns & 
69$\; \varepsilon_{\rm rec}$ \\
$\overline{\nu}_\mu\; e$ & \ \ $\mu^+$ DAR & 2197ns & 
99$\; \varepsilon_{\rm rec}$ \\
$\nu_e\; e$         & \ \ $\mu^+$ DAR & 2197ns & 
630$\; \varepsilon_{\rm rec}$ \\
\hline
\multicolumn{2}{|l}{Total $\nu\; e$ events \ \ \ $T_e >$ 20MeV}  &  &
{\bf 798}$\; \varepsilon_{\rm rec}$ \\
\hline\hline
$^{12}{\rm C}(\nu_e,e^-)^{12}{\rm N}_{\rm gs}$ & \ \ $\mu^+$ DAR & 2197ns &
4689$\; \varepsilon_{\rm rec}$ \\
$^{12}{\rm C}(\nu_e,e^-)^{12}{\rm N}^*$        & \ \ $\mu^+$ DAR & 2197ns & 
2147$\; \varepsilon_{\rm rec}$ \\
$^{13}{\rm C}(\nu_e,e^-)X$                     & \ \ $\mu^+$ DAR & 2197ns &
308$\; \varepsilon_{\rm rec}$ \\
\hline
\multicolumn{2}{|l}{Total $\nu_e\;{\rm C}$ events \ \ \ $T_e >$ 20MeV}  &  &
{\bf 7144}$\; \varepsilon_{\rm rec}$ \\
\hline\hline
$^{12}{\rm C}(\nu_\mu,\nu_\mu)^{12}{\rm C}^*_{15.11}$ & \ \ $\pi^+$ DAR & 26ns & 
1851$\; \varepsilon_{\rm rec}$ \\
 $^{12}{\rm C}(\overline{\nu}_\mu,\overline{\nu}_\mu)^{12}{\rm C}^*_{15.11}$ &
 \ \ $\mu^+$ DAR
& 2197ns & 3752$\; \varepsilon_{\rm rec}$ \\
$^{12}{\rm C}(\nu_e,\nu_e)^{12}{\rm C}^*_{15.11}$ & \ \ $\mu^+$ DAR & 2197ns  & 
3093$\; \varepsilon_{\rm rec}$ \\
\hline
Total $\nu_e\;{\rm C}$ NC events                     &  &  &
{\bf 8696}$\; \varepsilon_{\rm rec}$ \\
\hline\hline
$^{12}{\rm C}(\nu_\mu,\mu^-)X$ & \ \ $\pi^+$ DIF                  & 26ns & 
$\le 278\; \varepsilon_{\rm rec}$  \\
$^{12}{\rm C}(\overline{\nu}_\mu,\mu^+)X$ & \ \ $\pi^-$ DIF                  & 26ns & 
$\le 82\; \varepsilon_{\rm rec}$ \\
$p(\overline{\nu}_\mu,\mu^+)n$ & \ \ $\pi^-$ DIF                  & 26ns & 
272$\; \varepsilon_{\rm rec}$ \\
\hline
\end{tabular}
}
\end{table}

%
%
\begin{table}[htbp]
\caption{\label{tab:LSNDvSNS} Comparison of LSND and 250 ton SNS Detector
Both with Efficiency $\varepsilon=0.42$}

\vskip 0.2in
\begin{tabular}{ccc}
parameter & \ \ \ \ \ \  LSND (1993-98) \ \ \ \ \ \  &
                   \ \ \ \ \ \ SNS 250 ton detector \ \ \ \ \ \  \\
\hline
beam kinetic energy    & 798 MeV & 1300 MeV \\
$\nu/$p       & 0.079 & 0.135 \\
fiducial mass & 86.5 tons & 250 tons \\
free protons $N_t$           & $7.4\times 10^{30}$ & $4.3\times 10^{31}$ \\
$\nu$ flux                   & $1.26\times 10^{14}\nu/{\rm cm}^2$ & 
                               $6.45\times 10^{13}\nu/{\rm cm}^2/{\rm year}$ \\
$\nu$ flight path $L$        & 30m & 60m \\
Maximum mixing events ($\varepsilon=42\%$) & 33,300 & 49,500 / beam year \\
beam duty factor                  & $6\times 10^{-2}$ & $4.2\times 10^{-5}$ \\
\hline
\end{tabular}\end{table}

%
%
\begin{table}[htbp]
\caption{\label{tab:concl2} Oscillation Event Rates and Backgrounds in 250tons of Mineral Oil}

\vskip 0.2in
{\footnotesize
\begin{tabular}{|llr|}
\hline
\multicolumn{3}{|c|}{\bf DAR $\overline{\nu}_\mu\to\overline{\nu}_e$ 
Oscillation Signal Event Rates} \\
\hline
($\sin^2 2\theta$,$\Delta m^2$) & & Events \\
\hline
(0.0025,1 eV$^2$)   &  &
$266\; \varepsilon_{\rm rec}$ \\
(0.0075,0.5 eV$^2$) &  &
$592\; \varepsilon_{\rm rec}$ \\
(0.040,0.2eV$^2$) &  &
$693\; \varepsilon_{\rm rec}$ \\
\hline\hline
\multicolumn{3}{|c|}{\bf DAR $\overline{\nu}_\mu\to\overline{\nu}_e$ 
Oscillation Background Event Rates} \\
\hline
Intrinsic $\overline{\nu}_e$ & $8.5\times 10^{-4}$ per DAR $\nu$ &
81$\; \varepsilon_{\rm rec}$ \\
$\nu_\mu\; ^{12}{\rm C}$ & $\pi^+$ DIF                  & 
$<0.15\; \varepsilon_{\rm rec}$  \\
$\overline{\nu}_\mu\; ^{12}{\rm C}$ & $\pi^-$ DIF                  & 
$\le 0.3\; \varepsilon_{\rm rec}$  \\
$\overline{\nu}_\mu\; p$ & $\pi^-$ DIF                  & 
$4.4\; \varepsilon_{\rm rec}$ \\
mis-ID $\mu$, $\mu$ behind PMT & &
$<0.5\; \varepsilon_{\rm rec}$  \\
Beam Off ($R_\gamma >1$)  &     &  $3\; \varepsilon_{\rm rec}$  \\
\hline
\end{tabular}
}
\end{table}

%
%
\begin{table}[htbp]
\caption{\label{tab:other} Possible \v{C}erenkov Liquids.  Information from
Refs.~\protect\cite{CRC,ILO,PDGanpm,ESTAR}}

{\footnotesize
\begin{tabular}{lllrrrrrr}
         &       &          &         &           & Radiation & 50MeV $e^-$ & & \\
Physics  & Stuff & Chemical & Density & $n_D$\ \  & Length \ \ \ & Range \ \ \ & M.P. & B.P. \\
\hline
$\nu_e\;{\rm C}$ \ CC & mineral oil  &  ${\rm C_nH_{2n+2}}$
         &  860kg/m$^3$ &  1.47  & 44.8g/cm$^2$ & 14.8g/cm$^2$ & & $>150^\circ$C   \\
$\nu\;{\rm C}$ \ NC   &              &  $n \simeq 20$
         &             &  &  &            \multicolumn{3}{l}{add butyl-PBD for n's} \\
\hline
$\nu_e\;{\rm O}$ \ CC & water        &  ${\rm H_2 O}$
         & 1000kg/m$^3$  &  1.33  & 36.1g/cm$^2$ & 19.8g/cm$^2$ &  0$^\circ$C & 100$^\circ$C \\
$\nu\;{\rm O}$ \ NC   &              &
         &             &  &  &   \multicolumn{3}{l}{add Gd or NaCl to see n's?} \\
\hline
$\nu\;{\rm N}$      & ammonia      &  ${\rm N H_3}$
         &  771kg/m$^3$  & 1.33   & 41g/cm$^2$ & 16.8g/cm$^2$ &  & $-33^\circ$C \\

                     & ammonia+water &
         \multicolumn{3}{l}{(90g ${\rm N H_3}$ per 100g ${\rm H_2 O}$ at 10$^\circ$C)} &
               \multicolumn{3}{l}{ add ${\rm H_2 O}$ for room temp liquid} \\
\hline
$\nu\;{\rm Cl}$      & tetrachloroethene & ${\rm C_2 Cl_4}$
         & 1625kg/m$^3$  & 1.51   & 21g/cm$^2$ & 19.8g/cm$^2$ & $-19^\circ$C & 121$^\circ$C \\
\hline
$\nu\;{\rm Br}$      & tribromoethene    & ${\rm C_2 H Br_3}$
         & 2708kg/m$^3$  & 1.604  & 12.3g/cm$^2$ & 17.3g/cm$^2$ &  & 163$^\circ$C  \\
\hline
$\nu\;{\rm Br}$       & methylene bromide  & ${\rm C H_2 Br_2}$
         &  2497kg/m$^3$ & 1.542  & 12.2g/cm$^2$ & 15.7g/cm$^2$ & $-53^\circ$C & 97$^\circ$C \\
\hline
$\nu\;{\rm Br}$       & ethylene dibromide   & ${\rm C_2 H_4 Br_2}$
         &  2179kg/m$^3$ & 1.534  & 12.9g/cm$^2$ & 15.5g/cm$^2$ & 10$^\circ$C & 131$^\circ$C \\
\hline
$\nu\;{\rm I}$       & ethylene iodide  & ${\rm C_2 H_5 I_1}$
         & 1936kg/m$^3$ & 1.742  & 10.0g/cm$^2$ & 15.0g/cm$^2$ & $-108^\circ$C & 73$^\circ$C \\
\hline
$\nu\;{\rm I}$       & methylene iodide  & ${\rm C H_2 I_2}$
         & 3325kg/m$^3$  & 1.742  & 8.9g/cm$^2$ & 15.1g/cm$^2$ & 6.1$^\circ$C & 182$^\circ$C \\
\hline
$\nu\;{\rm D}$       & deuterated water  & ${\rm D_2O}$
         & 1105kg/m$^3$  & 1.34   & &  \multicolumn{3}{l}{ transparent sub-volume} \\
\hline
\end{tabular}
}
\end{table}

%
%
\begin{table}[htbp]
\caption{\label{tab:nuI} Event rates in a methylene iodide filled detector}

\vskip 0.2in
\begin{tabular}{|ccc|}
\hline
Process  & $\sigma$ & Events per beam year \\
\hline
\ \ $^{127}{\rm I}(\nu_e,e^-)^{127}{\rm Xe}_{\rm bound\ states}$ \ \ & 
\ \ $2.75\times 10^{-40}{\rm cm}^2$ \ \ & {\bf 77,000}$\; \varepsilon_{\rm rec}$ \\
\ \ $^{127}{\rm I}(\nu_e,e^-)^{127}{\rm Xe}^*_{\rm total}$ \ \ &  &
 {\bf 131,000}$\; \varepsilon_{\rm rec}$ \\
\ \ $^{127}{\rm I}(\nu_e,e^-) X\; n$ \ \ & ? & ? \\
\ \ $^{127}{\rm I}(\nu,\nu^\prime)^{127}{\rm I}^*$ \ \ & ? & ? \\
\hline
$^{12}{\rm C}(\nu_e,e^-)^{12}{\rm N}_{\rm gs}$ & $6.76\times 10^{-42}{\rm cm}^2$ &
 946$\; \varepsilon_{\rm rec}$ \\
$^{12}{\rm C}(\nu_e,e^-)^{12}{\rm N}^*$         & $5.41\times 10^{-42}{\rm cm}^2$ &
 602$\; \varepsilon_{\rm rec}$ \\
$^{13}{\rm C}(\nu_e,e^-)X$                     & $50\times 10^{-42}{\rm cm}^2$   &
 78$\; \varepsilon_{\rm rec}$ \\
\hline
${\rm C}(\nu_e,e^-)X$ \ total                  & &
 {\bf 1626$\; \varepsilon_{\rm rec}$} \\
\hline
\end{tabular}
\end{table}

\newpage

\end{document}